# Electronic properties and quasiparticle model of monolayer MoSi$_2$N$_4$

Zhenwei Wang,[1,3] Xueheng Kuang,[1] Guodong Yu,[1] Peiliang Zhao,[1] Hongxia Zhong,[2,1,*] and Shengjun Yuan[1,†]

[1]*School of Physics and Technology, Wuhan University, Wuhan 430072, China*
[2]*School of Mathematics and Physics, China University of Geosciences (Wuhan), Wuhan 430074, China*
[3]*Department of Physics, Anhui Normal University, Wuhu 241000, China*

In this paper, we theoretically investigated the electronic properties of monolayer MoSi$_2$N$_4$ by combining first-principles calculations and symmetry analyses. Spin-orbital coupling resulted in band splitting, whereas a horizontal mirror symmetry constrained the spin polarization to be along the $z$ direction. In addition, a three-band tight-binding model was constructed to describe the low-energy quasiparticle states of monolayer MoSi$_2$N$_4$, which can be generalized to strained MoSi$_2$N$_4$ and its derivatives. The calculations using the tight-binding model showed an undamped $\sqrt{q}$-dependent plasmon mode, consistent with the results of first-principles calculations. The developed model is suitable for future theoretical and numerical investigations of low-energy properties in MoSi$_2$N$_4$ family materials. Furthermore, the study of the electronic properties of monolayer MoSi$_2$N$_4$ paves a way for its applications in spintronics and plasmonics.

## I. INTRODUCTION

Since the discovery of graphene [1], more novel two-dimensional (2D) materials are being synthesized constantly [2–7]. In the monolayer limit, 2D materials exhibit various unique properties, which have attracted considerable interest in both theoretical and experimental studies [8–10]. For example, the discovery of 2D transition-metal dichalcogenides provided the impetus for spintronics [11,12]. In these materials, strong spin-orbit coupling (SOC) is crucial for realizing spin and valley polarization. However, SOC is detrimental to the spin lifetime, which greatly limits the performance of potential spintronic devices [13]. Persistent spin polarization is proposed as a solution to the poor spin lifetime, which has been demonstrated experimentally in many 2D materials [14–16]. Thus, the discoveries of new 2D materials exhibiting persistent spin polarization are very important for spintronic applications [17,18]. In addition to spintronics, huge attention has also been attracted to the research of plasmonics [19,20]. Plasmons can be viewed as the quantized collective motion of surface electrons originating from the coupling of photon and free-electron gases. They possess many attractive properties, such as a strong interaction with light and high optical enhancement, which currently drive the development of novel devices for ultrasensitive detection [21], improved photovoltaics [22], nanoscale photometry [23], cancer therapy [24], and nonlinear optics [25].

Recently, monolayer MoSi$_2$N$_4$ has been synthesized via chemical vapor deposition (CVD) [26,27]. The excellent mechanical properties and ambient stability of monolayer MoSi$_2$N$_4$ have triggered a competition among researchers to explore its electronic, phononic, and transport properties [28–31]. However, the spin polarization and plasmons of monolayer MoSi$_2$N$_4$ still need to be studied. Besides first-principles calculations, effective models provide alternative approaches to investigate electronic structures. Up to now, various effective models have been proved to capture relevant electronic states in 2D materials, such as graphene [32,33] and transition-metal dichalcogenides [34,35]. An effective two-band $k \cdot p$ model of monolayer MoSi$_2$N$_4$ has been constructed recently [29], whereas the tight-binding (TB) model of monolayer MoSi$_2$N$_4$ has yet to be established. TB models can cover a much larger momentum space and energy range, and are useful for studying various physical properties of 2D materials. Therefore, constructing a TB model is important for describing the low-energy quasiparticle states of monolayer MoSi$_2$N$_4$.

In this paper, we first investigate the spin polarization and plasmons of monolayer MoSi$_2$N$_4$ based on first-principles calculations and effective model analyses. The broken inversion symmetry and SOC result in band spin splitting, and horizontal mirror symmetry $M_z$ constrains spin polarization to be along the $z$ direction. This means that charge carriers have a long spin lifetime in monolayer MoSi$_2$N$_4$, which is crucial for spintronic applications [13]. Then, we developed a three-band TB model to describe the low-energy electronic states of monolayer MoSi$_2$N$_4$. The model can be generalized to strained MoSi$_2$N$_4$ and its derivatives. Based on this model, we calculated the plasmons of hole-doped MoSi$_2$N$_4$, and the results agree well with those of the first-principles calculations. Therefore, the developed TB model can effectively capture the relevant low-energy electronic states of monolayer MoSi$_2$N$_4$. This is important for future theoretical and numerical studies on MoSi$_2$N$_4$ family materials [36].

This paper is organized as follows: Section II presents the crystal structure and calculation methodology. Section III discusses the spin polarization from the perspective of symmetry. A tractable three-band TB model of monolayer MoSi$_2$N$_4$ is deduced in Sec. IV. Based on this model, hole carrier-induced plasmon dispersion spectra of monolayer MoSi$_2$N$_4$ are presented in Sec. V. The conclusion is given at the end of this paper.

---

*Corresponding author: hxzhong@whu.edu.cn
†Corresponding author: s.yuan@whu.edu.cn

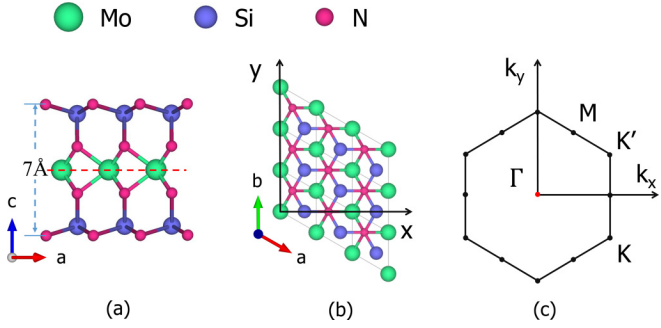

FIG. 1. (a) Side and (b) top views of monolayer MoSi$_2$N$_4$. The red dashed line in (a) indicates the horizontal mirror plane. The diamond lattice in (b) is the primitive cell with lattice constant $a$. First Brillouin zone of monolayer MoSi$_2$N$_4$ is shown in (c).

## II. CRYSTAL STRUCTURE AND COMPUTATIONAL METHOD

Monolayer MoSi$_2$N$_4$ belongs to the hexagonal lattice with space group $P\bar{6}m2$ (No. 187). Along the $z$ direction, it has a septuple atomic layer with 7 Å thickness. It has mirror symmetry with a layer of Mo atoms in the middle. On both sides of the Mo atoms, there are two layers of N atoms and one layer of Si atoms. The top and side views of the MoSi$_2$N$_4$ supercell crystal structure are shown in Figs. 1(a) and 1(b). The corresponding first Brillouin zone (BZ) of monolayer MoSi$_2$N$_4$ is shown in Fig. 1(c). All calculations herein are implemented in the Vienna *ab initio* simulation package (VASP) [37,38], which is based on density functional theory (DFT) [39,40]. The generalized gradient approximation (GGA) with the Perdew-Burke-Ernzerhof (PBE) functional and projector augmented-wave (PAW) pseudopotential [41] are employed in our calculations. The cutoff energy is set to 500 eV, and the maximum force is set to 0.01 eV. A $9 \times 9 \times 1$ Monkhorst-Pack $k$-point grid [42] is used for structure optimization, and a $40 \times 40 \times 1$ grid is used for self-consistent calculations. A vacuum layer with a thickness of 30 Å is adopted to avoid hypothetical interactions between neighboring cells. Based on these parameters, the optimized lattice constants of monolayer MoSi$_2$N$_4$ are $a = b = 2.911$ Å, which agree well with the value (2.909 Å) reported in Ref. [29]. The Wannier functions and symmetrized TB Hamiltonian are constructed using maximally localized Wannier functions [43,44]. The plasmon spectra of monolayer MoSi$_2$N$_4$ are calculated by using the Wannier TB Hamiltonian and the random phase approximation method.

## III. SYMMETRY AND SPIN POLARIZATION

The band structure of monolayer MoSi$_2$N$_4$ is shown in Fig. 2(a). It has a semiconductor band structure with an indirect band gap of 1.786 eV. The trigonal crystal field splits five Mo $d$ orbitals into three categories: $A'_1$ $\{d_{z^2}\}$, $E'$ $\{d_{xy}, d_{x^2-y^2}\}$, and $E''$ $\{d_{xz}, d_{yz}\}$. They correspond to the irreducible representations of the $D_{3h}$ point group at the $\Gamma$ point: $\Gamma_1$, $\Gamma_5$, and $\Gamma_6$, respectively. This indicates that the low-energy quasiparticle states of monolayer MoSi$_2$N$_4$ are mainly dominated by $3d$ orbitals, which are important to understand the DFT results and interactions between orbitals. When SOC is included, the bands along $\Gamma$-M remain spin degenerate but others split, as shown in Fig. 2(b). The splitting strength of the first valence band over the BZ is shown in Fig. 2(c). The strongest band splitting (0.149 eV) occurs at the K and K' points. Splitting gradually decreases as the $k$ point moves away from the K and K' points and drops to zero on the $\Gamma$-M high-symmetry line. Our calculations show that the spin polarizations of all band eigenstates are along the $z$ direction. Figure 2(d) shows the energy isosurface for a binding energy 0.5 eV [marked by a red dotted line in Fig. 2(b)]. The orientations of the spin polarization $S_z$ on the isosurface are indicated using different colors. The results show a spin-valley locked band structure, which is consistent with a previous report [30].

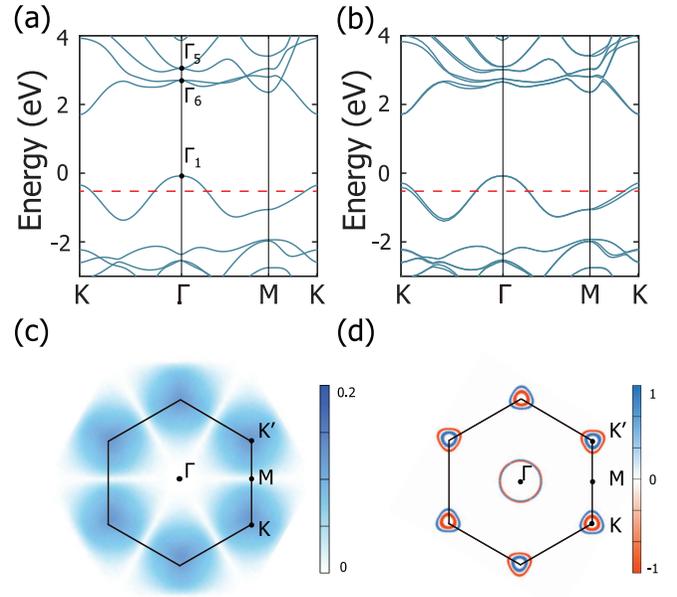

FIG. 2. Band structures of monolayer MoSi$_2$N$_4$ (a) without SOC and (b) with SOC. The red line represents the energy level for binding energy 0.5 eV. (c) Band splitting strength for the first valence band in the entire BZ. The color bar denotes the value of the splitting intensity. (d) Spin polarization $S_z$ projected on the energy isosurface for a binding energy 0.5 eV. Blue and red colors denote the spin-up and spin-down states, respectively.

The phenomenon that all the states are spin polarized only along the $z$ direction can be explained from the aspect of symmetry. Due to the horizontal mirror symmetry $M_z$, the Hamiltonian commutes with $M_z$. The eigenvalues ($\Lambda$) of $M_z$ are good quantum numbers, and all the eigenstates ($|\psi_{nk}\rangle$) of this system can be labeled using eigenvalues of $M_z$, namely, $M_z|\psi_{nk}\rangle = \Lambda|\psi_{nk}\rangle$. In the 1/2-spin space, $M_z$ and $C_{2z}$ share the same matrix [45], where $C_{2z}$ is the rotation around the $z$ axis by $180°$. Therefore, the mirror symmetry $M_z$ can be represented as $D^{1/2}(\pi, 0, 0) = -i\sigma_z$, where $D^{1/2}(\pi, 0, 0)$ is an element of the SU(2) group, and $\sigma_z$ is the $z$ component of the Pauli matrix. The eigenvalue of $M_z$ can be either $i$ or $-i$, which classifies the Bloch states into two groups. For a given state $|\psi_{nk}\rangle$ with any eigenvalue of $M_z$, the anticommutation relationship $M_z \sigma_{x,y} = -\sigma_{x,y} M_z$ results in [46]

$$\begin{aligned}\langle \sigma_{x,y}\rangle &= \langle \psi_{nk}|\sigma_{x,y}|\psi_{nk}\rangle \\ &= \langle \psi_{nk}|M_z^{-1}\sigma_{x,y}M_z|\psi_{nk}\rangle \\ &= -\langle \psi_{nk}|\sigma_{x,y}|\psi_{nk}\rangle = 0.\end{aligned} \quad (1)$$

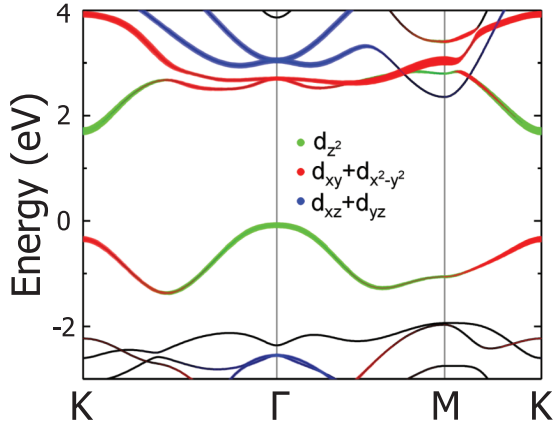

FIG. 3. Orbital projected band structure for monolayer MoSi$_2$N$_4$. Symbol size is proportional to the weight projected onto the particular orbital of the Mo atom.

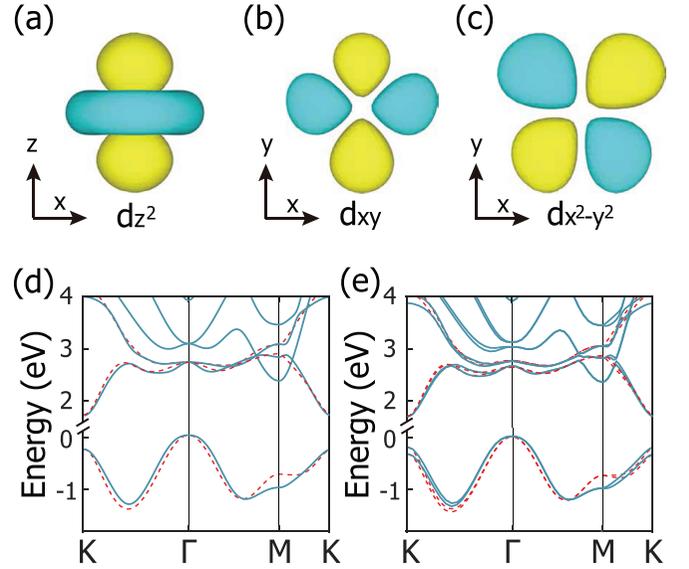

FIG. 4. (a) $d_{z^2}$-, (b) $d_{xy}$-, and (c) $d_{x^2-y^2}$-like Wannier orbitals of monolayer MoSi$_2$N$_4$ corresponding to the basis of the TB Hamiltonian. Band structures of monolayer MoSi$_2$N$_4$ (d) without and (e) with SOC. The red (blue) lines correspond to the TB (DFT) results.

Accordingly, the horizontal mirror symmetry $M_z$ eliminates spin polarization in the $x$-$y$ plane preserving the spin polarization along the $z$ direction.

For any $\mathbf{k}$ point located on the line $\Gamma$-M along the $k_x$ axis, the Hamiltonian $H(\mathbf{k})$ exhibits a point group symmetry of $C_{2v}(=C_{2x}$ and $M_z)$, where $C_{2x}$ is the rotation around the $x$ axis by 180°. Under these constraints of symmetry, the Hamiltonian satisfies the following equation,

$$\begin{aligned}M_z H(\mathbf{k}) M_z^{-1} &= H(k_x, k_y), \\ C_{2x} H(\mathbf{k}) C_{2x}^{-1} &= H(k_x, -k_y),\end{aligned} \quad (2)$$

and an effective two-band Hamiltonian up to second-order correction of $\mathbf{k}$ can be expressed as

$$H(\mathbf{k}) = m_1 k_x + m_2 k_x^2 + m_3 k_y^2 \\ + (m_4 k_y + m_5 k_x k_y)\sigma_z, \quad (3)$$

where $m_{i=1-5}$ can be chosen as real. For any $\mathbf{k}$ with $k_y = 0$, the Hamiltonian $H(\mathbf{k})$ does not contain the spin operator. Hence, its eigenvalues are always spin degenerate, and the bands on the line $\Gamma$-M do not undergo spin splitting.

## IV. THREE-BAND TB MODEL

Figure 3 shows the orbital projected band structure of monolayer MoSi$_2$N$_4$. The highest occupied band and lowest two unoccupied bands are dominated by $d_{z^2}$, $d_{xy}$, and $d_{x^2-y^2}$ orbitals, which disentangle $d_{xz}$ and $d_{yz}$ orbitals. Therefore, we can choose $d_{z^2}$, $d_{xy}$, and $d_{x^2-y^2}$ orbitals as the initial Wannier bases to establish a three-band TB model. The parametrization procedure is based on maximally localized Wannier functions (MLWFs), avoiding the disentanglement procedure and guaranteeing the WFs uniquely defined within the scheme of maximal localization. The real-space distribution of MLWFs is shown in Fig. 4. Three $d$-like orbitals are localized on the Mo atom, resulting in three MLWFs per cell.

The nonrelativistic TB model is given by a 3 × 3 effective Hamiltonian. The matrix elements of the Hamiltonian in reciprocal space are given by

$$H_{mn}(\mathbf{k}) = \langle m, \mathbf{k}|H|n, \mathbf{k}\rangle = t_{mn}(\mathbf{r})\sum_{\mathbf{r}} e^{i\mathbf{k}\cdot\mathbf{r}}, \quad (4)$$

where $m$ and $n$ correspond to the $d_{z^2}$-, $d_{xy}$-, and $d_{x^2-y^2}$-like orbitals. For simplicity, these orbitals are denoted by numbers 1, 2, and 3, respectively. $t_{mn}$ is the effective hopping parameter describing the interaction between different orbitals, and $\mathbf{r}$ is the corresponding hopping vector. To make the model more tractable and sufficiently accurate, we first discard the hopping term with an interatomic distance longer than the third neighbor. Next, the residual hopping parameters are further optimized by ignoring the hopping terms with amplitudes $|t| < 10$ meV. The optimized hopping parameters are summarized in Table I, and the on-site potential energies of three orbitals are listed in Table II. In reciprocal space, we obtain the Hamiltonian matrix using a Fourier transform, and the results are shown in the Appendix.

TABLE I. Hopping amplitudes (in eV) of monolayer MoSi$_2$N$_4$ assigned to Eq. (4). $\alpha$, $\beta$, and $\gamma$ denote the nearest, secondary, and third neighbors of the Mo atom sites. $i$ is the corresponding ordering of $t_{11}$, $t_{22}$, $t_{33}$, $t_{12}$, $t_{23}$, $t_{13}$, respectively.

| $i$ | 1 | 2 | 3 | 4 | 5 | 6 |
| --- | --- | --- | --- | --- | --- | --- |
| $\alpha$ | −0.220 | −0.180 | −0.050 | −0.500 | −0.210 | −0.680 |
| $\beta$ | 0.070 | 0.020 | 0.010 | −0.040 | −0.010 | −0.020 |
| $\gamma$ | 0.020 | 0.050 | 0.010 | −0.030 | −0.040 | −0.060 |

TABLE II. Original Hamiltonian parameters (in eV) of MoSi$_2$N$_4$ and correctional parameters of strained MoSi$_2$N$_4$ and WSi$_2$N$_4$. $V_{d_{z^2}}$, $V_{d_{xy}}$, and $V_{d_{x^2-y^2}}$ are the chemical potentials of three atomlike orbitals. $\alpha_1$ is the nearest-neighbor hopping between $d_{z^2}$ orbitals, and $\lambda$ is the strength of SOC.

| | $V_{d_{z^2}}$ | $V_{d_{xy}}$ | $V_{d_{x^2-y^2}}$ | $\alpha_1$ | $\lambda$ |
|---|---|---|---|---|---|
| MoSi$_2$N$_4$ | 0.780 | 2.080 | 2.080 | −0.220 | 0.064 |
| MoSi$_2$N$_4$ (−2%) | 0.950 | 2.120 | 2.120 | −0.260 | 0.064 |
| WSi$_2$N$_4$ | 0.920 | 2.050 | 2.050 | −0.250 | 0.180 |

Figure 4(d) compares the DFT and TB band structures. The two structures agree well with each other, except for the deviation around the M point. This deviation is because the three-band TB model ignores the $p$ orbitals of Si atoms, which mainly contribute to the electronic states around the M valley. Nevertheless, the TB model is sufficient to describe the physics of the valence band maximum (VBM) at $\Gamma$ and conduction band minimum (CBM) at the K point. These features also appear in monolayer transition-metal dichalcogenides [34]. Thus, the proposed three-band TB model can be effectively used to describe the electronic and transport properties in the low-energy region of monolayer MoSi$_2$N$_4$.

Next, we focus on the SOC in monolayer MoSi$_2$N$_4$. Although the Wannier orbitals are not exactly the same as the real 3$d$ atomic orbitals, there is an extreme similarity between them (seen as 3$d$ atomic orbitals when SOC is induced). Under the atomic orbital basis group $\{d_{z^2}\uparrow, d_{xy}\uparrow, d_{x^2-y^2}\uparrow, d_{z^2}\downarrow, d_{xy}\downarrow, d_{x^2-y^2}\downarrow\}$, the SOC contribution to the Hamiltonian is expressed as [34]

$$H' = \lambda \mathbf{L}\cdot\mathbf{S} = \frac{\lambda}{2}\begin{pmatrix} L_z & 0 \\ 0 & -L_z \end{pmatrix}, \quad (5)$$

in which

$$L_z = \begin{pmatrix} 0 & 0 & 0 \\ 0 & 0 & 2i \\ 0 & -2i & 0 \end{pmatrix} \quad (6)$$

is the $z$ component of the orbital angular momentum matrix in the above-mentioned three $d$ orbital bases. $\lambda$ characterizes the strength of SOC. Next, the TB+SOC Hamiltonian is expressed as follows,

$$H_{\text{soc}}(\mathbf{k}) = I_2 \otimes H(\mathbf{k}) + H', \quad (7)$$

where $I_2$ is the 2 × 2 identity matrix. By fitting the DFT results with SOC, we obtain a $\lambda$ value of 0.064. The band structures from both the DFT and TB Hamiltonians with SOC are shown in Fig. 4(e). The TB+SOC model matches well with the DFT+SOC bands, confirming the validity of the SOC Hamiltonian.

During the synthesis of MoSi$_2$N$_4$ samples, their electronic properties can be efficiently tuned by lattice strains [26]. Thus, the strain effect should be considered in the proposed TB model. Based on the obtained Hamiltonian of unstrained MoSi$_2$N$_4$, the on-site energies determine the band gap, and the $\alpha_1$ hopping term is mainly responsible for adjusting the band edges. Thus, by fitting the three on-site energies and $\alpha_1$ parameters listed in Table II, we get the TB band structures at a small biaxial strain of −2%, as shown in Fig. 5. With compressive biaxial strain, $V_{d_{z^2}}$ is significantly enhanced, resulting in a large band gap. Moreover, the $\alpha_1$ hopping term controls the VBM shift. The larger the compressive strain, the stronger is the $V_{d_{z^2}}$ orbital interaction. Accordingly, the $\alpha_1$ amplitude is enhanced, decreasing $\Gamma_1$. When SOC is included in strained MoSi$_2$N$_4$, the SOC strength $\lambda$ remains invariable.

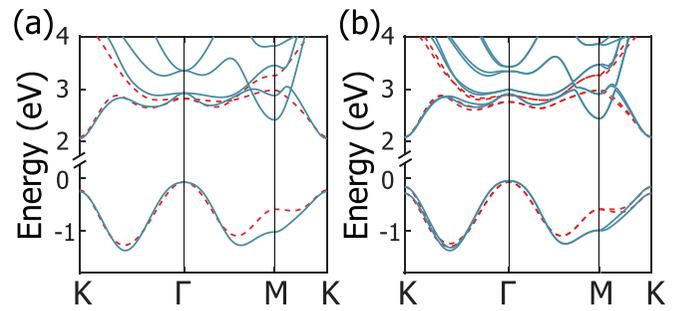

FIG. 5. Band structures of monolayer MoSi$_2$N$_4$ (a) without and (b) with SOC at a biaxial strain of −2%. The red (blue) lines correspond to the TB (DFT) results.

Finally, by modifying the leading hopping terms, we extend the TB model of monolayer MoSi$_2$N$_4$ to WSi$_2$N$_4$ (see Fig. 6). The optimized lattice constant is 2.914 Å, and the indirect band gap is 2.130 eV, which is larger than the band gap of monolayer MoSi$_2$N$_4$. The larger band gap is attributed to the higher on-site potential energy $V_{d_{z^2}}$ in WSi$_2$N$_4$ (0.920 eV) compared with that of MoSi$_2$N$_4$ (0.780 eV), as shown in Table II. Compared with MoSi$_2$N$_4$, SOC splitting is enhanced in WSi$_2$N$_4$. This is because the W atom is heavier than the Mo atom. The enhanced SOC splitting is 0.405 eV at the K valley, which is determined by the larger SOC strength $\lambda = 0.180$. Adjusting the potential energy, hopping term $\alpha_1$, and SOC strength $\lambda$ in the TB model of monolayer MoSi$_2$N$_4$, we can obtain the Hamiltonians of MoSi$_2$N$_4$ derivatives. This is important for studying the electronic properties of MoSi$_2$N$_4$ family materials [36].

## V. PLASMON AND DYNAMICAL POLARIZATION

Herein, the electronic excitation (plasmon) is investigated in a hole-doped monolayer MoSi$_2$N$_4$. The dynamical

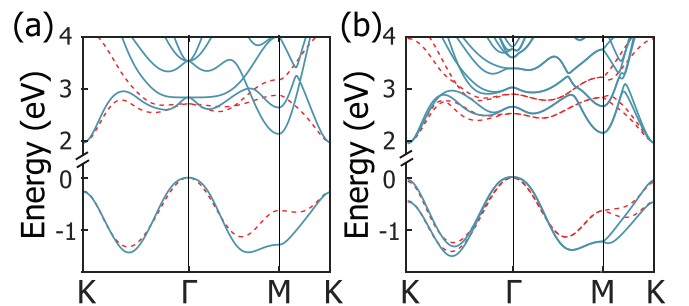

FIG. 6. Band structures of monolayer WSi$_2$N$_4$ (a) without and (b) with SOC. The red (blue) lines correspond to the TB (DFT) results.

polarization $\Pi(\mathbf{q}, \omega)$ is calculated using the Lindhard function

$$\Pi(\mathbf{q}, \omega) = -\frac{g_s}{(2\pi)^2} \int_{BZ} d^2\mathbf{k} \sum_{l,l'} \frac{n_F(E_{\mathbf{k}l}) - n_F(E_{\mathbf{k}'l'})}{E_{\mathbf{k}l} - E_{\mathbf{k}'l'} + \hbar\omega + i\delta} \quad (8)$$
$$\times |\langle \mathbf{k}'l' | e^{i\mathbf{q}\cdot\mathbf{r}} | \mathbf{k}l \rangle|^2,$$

where $l$ is the band index, $\mathbf{k}' = \mathbf{k} + \mathbf{q}$, $n_F(E) = (e^{\beta(E-\mu)} + 1)^{-1}$ is the Fermi-Dirac distribution for chemical potential $\mu$, and $|\mathbf{k}l\rangle$ and $E_{\mathbf{k}l}$ are the eigenstates and eigenvalues, respectively. Herein, $\mu$ is set to $-0.5$ eV, corresponding to the hole carrier concentration of $2.7 \times 10^{-13}$ cm$^{-2}$. The charge carriers can be injected via electrical gating [47,48] or chemical doping [49,50]. The integral is calculated in the first BZ, including both inter- and intraband transitions. After considering the electron-electron interaction $V(q) = \frac{2\pi e^2}{\varepsilon_B q}$, the dielectric function is calculated by using dynamical polarization as follows:

$$\epsilon(\mathbf{q}, \omega) = 1 - V(q)\Pi(\mathbf{q}, \omega). \quad (9)$$

Since the properties of monolayer $MoSi_2N_4$ are measured on a $SiO_2$ substrate [26], we take the dielectric constant $\varepsilon_B = 3.9$ to represent the environment of the silica substrate. Meanwhile, the long-wavelength limit of the dielectric function ($q = 0$) is not considered because it is irrelevant to the plasmon dispersion. The energy loss function related to the dielectric function can be expressed as follows:

$$S(\mathbf{q}, \omega) = -\text{Im}(1/\epsilon). \quad (10)$$

Here, we focus on the low-energy region of the excitation spectrum, which is relevant for future experimental probes and possible applications [51,52].

Considering that the intensities and frequencies of plasmons depend on the propagation direction, we calculate the plasmon spectra along the $\Gamma$-M and $\Gamma$-K high-symmetry directions with a length of 8.0 nm$^{-1}$. Figure 7(a) shows the plasmon spectrum of hole-doped $MoSi_2N_4$ at low energies without SOC. Around the $\Gamma$ point, a well-defined intraband plasmon branch is obtained. It exhibits a standard 2D electron gas $w_{pl} \propto \sqrt{q}$ dispersion, where $w_{pl}$ is the plasmon energy. This plasmon branch quickly damps into uncoupled electron-hole pairs at large $\mathbf{q}$. In regions away from the $\Gamma$ point, the plasmon dispersions exhibit anisotropy along different directions. This anisotropy is attributed to the anisotropic band dispersion for the $k$ point near the K and M points.

After taking SOC into account, the plasmon spectrum is shown in Fig. 7(b). Compared with the case without SOC, the plasmon branch has a larger slope in the small wave vector. This is attributed to the aforementioned spin splitting of the band structure. Although the splitting results in an interband transition, the energy of the band splitting is so small that the interband modes hardly interact with the intraband plasmons. Thus, the plasmons are continuous, which is different from the gapped plasmons in hole-doped $SnSe_2$ [53]. When we slightly adjust the doping chemical potential $\mu$, the plasmons are almost unchanged, which is different from the plasmons in monolayer antimonene [54]. This is because SOC causes only band splitting and barely affects the collective state density of electrons. These anisotropic plasmon spectra can be obtained experimentally via electron energy loss spectroscopy [55].

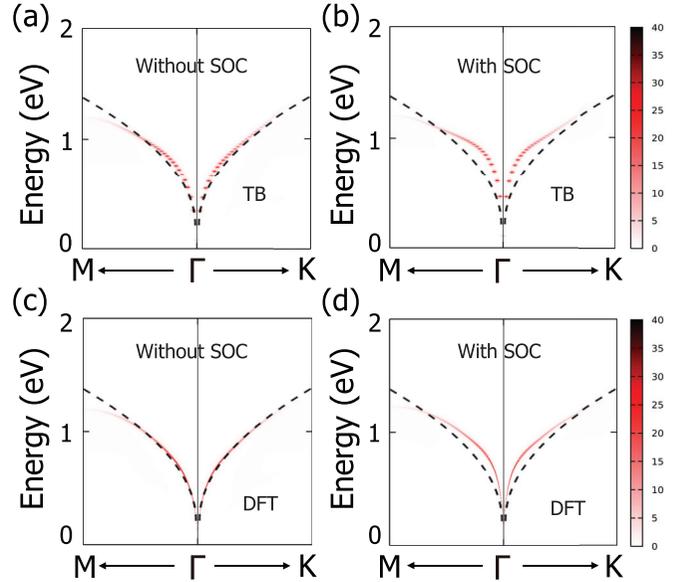

FIG. 7. Intensity of loss function $S(\mathbf{q}, \omega)$ in hole-doped $MoSi_2N_4$. (a) and (b) are the plasmon dispersions calculated using TB. (c) and (d) are the plasmon dispersions calculated using DFT. They are both along the $\Gamma$-M and $\Gamma$-K high-symmetry directions with length 8.0 nm$^{-1}$. The black dashed lines represent the plasmon $w_{pl} \propto \sqrt{q}$ dispersion.

To verify the reliability of the above-mentioned plasmon dispersion, we calculate the plasmon spectra of hole-doped $MoSi_2N_4$ using the DFT method, as shown in Figs. 7(c) and 7(d). The results agree well with those obtained using the proposed three-band TB model [see Figs. 7(a) and 7(b)]. This shows that the proposed TB model can sufficiently describe frequency-dependent plasmon dispersion and anisotropy.

## VI. CONCLUSION

In summary, we investigated the spin polarization of monolayer $MoSi_2N_4$ through first-principles calculations and symmetry analyses. Broken inversion symmetry and SOC result in band spin splitting, whereas the crystal symmetry $M_z$ constrains spin polarization along the $z$ direction. Further, we established a three-band TB model to describe the low-energy quasiparticle states of monolayer $MoSi_2N_4$, which can be extended to strained $MoSi_2N_4$ and its derivatives. Based on the developed TB model, the plasmon spectra of hole-doped monolayer $MoSi_2N_4$ were calculated, and the results agree well with those of the first-principles calculations. We not only predicted the spin polarization and plasmon properties of monolayer $MoSi_2N_4$ but also provided an effective model for future theoretical and numerical studies on $MoSi_2N_4$ family materials.

## ACKNOWLEDGMENTS

This work is supported by the National Key R&D Program of China (Grant No. 2018YFA0305800) and National Natural Science Foundation of China (Grants No. 11947218 and No. 12104421). Numerical calculations presented in this paper were performed on the supercomputing system in the Supercomputing Center of Wuhan University.

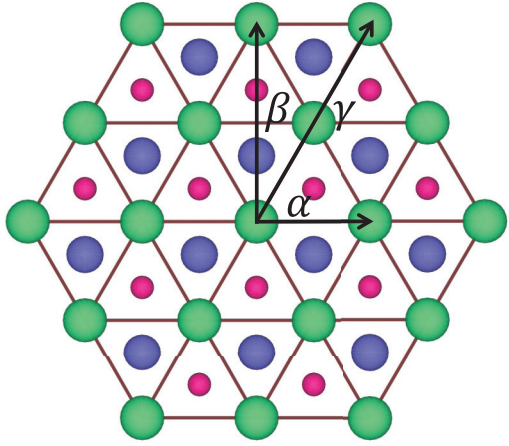

FIG. 8. Crystal structure of MoSi$_2$N$_4$ supercell. $\alpha$, $\beta$, and $\gamma$ denote the nearest, secondary, and third neighbor between the Mo atom sites.

## APPENDIX: DETAILS OF THE THREE-BAND MODEL

To clarify the TB model of monolayer MoSi$_2$N$_4$, we plotted the crystal structure of a MoSi$_2$N$_4$ supercell to explicate the neighboring Mo atom sites, as shown in Fig. 8. The nearest, second, and third neighbor are represented by the length of black arrows. The $3 \times 3$ TB Hamiltonian in reciprocal space is expressed as follows,

$$H(\boldsymbol{k}) = \begin{pmatrix} h_{11} & h_{12} & h_{13} \\ & h_{22} & h_{23} \\ \dagger & & h_{33} \end{pmatrix}, \quad (A1)$$

where the matrix elements are

$$h_{11} = V_{d_{z^2}} + 2\alpha_1 \cos(ak_x) + 2\gamma_1 \cos(2ak_x)$$
$$+ 4\cos\left(\frac{1}{2}ak_x\right)[\alpha_1 - \beta_1 + 2\beta_1 \cos(ak_x)]$$
$$\times \cos\left(\frac{\sqrt{3}}{2}ak_y\right) + 2[\beta_1 + 2\gamma_1 \cos(ak_x)]\cos(\sqrt{3}ak_y), \quad (A2)$$

$$h_{22} = V_{d_{xy}} + 2\alpha_2 \cos(ak_x) - 2\gamma_3 \cos(2ak_x)$$
$$- 4\cos\left(\frac{1}{2}ak_x\right)[\alpha_2 + \beta_2 - 2\beta_2 \cos(ak_x)]$$
$$\times \cos\left(\frac{\sqrt{3}}{2}ak_y\right) + 4[\beta_2 + \gamma_2 \cos(ak_x)]\cos(\sqrt{3}ak_y), \quad (A3)$$

$$h_{33} = V_{d_{x^2-y^2}} - 12\alpha_3 \cos(ak_x) - 2\gamma_6 \cos(2ak_x)$$
$$+ 4\cos\left(\frac{1}{2}ak_x\right)[\alpha_3 - 3\beta_3 + 6\beta_3 \cos(ak_x)]$$
$$\times \cos\left(\frac{\sqrt{3}}{2}ak_y\right) + 2[\beta_3 + 2\gamma_3 \cos(ak_x)]\cos(\sqrt{3}ak_y), \quad (A4)$$

$$h_{12} = -2\alpha_4 \cos(ak_x) + 2\gamma_6 \cos(2ak_x) + 2\cos(ak_x/2)$$
$$\times [\alpha_6 + \beta_4 - \alpha_2 - \beta_5 + 2(\beta_5 - \beta_4)\cos(ak_x)]$$
$$\times \cos(\sqrt{3}/2ak_y) + [\beta_4 - 2(\gamma_4 + \gamma_5)\cos(ak_x)]$$
$$\times \cos(\sqrt{3}ak_y) + 2i\cos(ak_x/2)[\alpha_2 + \alpha_6 + \beta_1 + \beta_2$$
$$+ 2(\beta_1 + \beta_2)\cos(ak_x)]\sin(\sqrt{3}/2ak_y)$$
$$- i[3\beta_4 + 2(\gamma_5 - \gamma_4)\cos(ak_x)]\sin(\sqrt{3}ak_y), \quad (A5)$$

$$h_{13} = 2i[-\alpha_2 + \alpha_6 + \beta_1 + \beta_2 + 2(\beta_1 + \beta_2)\cos(ak_x)]$$
$$\times \cos(\sqrt{3}/2ak_y)\sin(ak_x/2) - 2[-\alpha_2 - \alpha_6 + \beta_1$$
$$+ \beta_6 + 2(\beta_2 + \beta_6)\cos(ak_x)]\sin(ak_x/2)$$
$$\times \sin(\sqrt{3}/2ak_y) + 2i\sin(ak_x)[\alpha_4 + 2\gamma_3 \cos(ak_x)$$
$$+ 2i\gamma_6 \sin(\sqrt{3}ak_y)], \quad (A6)$$

$$h_{23} = -4i\sin(ak_x/2)[-2\alpha_5 \cos(\sqrt{3}/2ak_y) + i[\alpha_5$$
$$+ \beta_5 + 2\beta_5 \cos(ak_x)]\sin(\sqrt{3}/2ak_x) + \cos(ak_x/2)$$
$$\times [2\alpha_5 + \gamma_3 \cos(ak_x)] + i(\gamma_4 + \gamma_5)\sin(\sqrt{3}ak_y)]]. \quad (A7)$$